\documentclass[journal=jacsat,manuscript=letter,layout=twocolumn]{achemso} 

\usepackage[version=3]{mhchem} 
\usepackage[hidelinks]{hyperref}
\usepackage{orcidlink}
\usepackage{bm}
\usepackage{textcomp}
\def\qe{\textsc{Quantum ESPRESSO}\texttrademark}

\usepackage[normalem]{ulem}
\usepackage{xcolor}
\definecolor{amber}{rgb}{1,0.49,0}
\definecolor{darkgreen}{rgb}{0,0.55,0}
\definecolor{tangerine}{rgb}{0.944,0.522,0}
\definecolor{verde}{rgb}{0.,0.6,0}
\definecolor{rosso}{rgb}{0.9,0.0,0.2}
\definecolor{magenta}{rgb}{0.9,0.2,0.9}

\newif\ifhighlight

\newcommand{\highlight}{\highlighttrue}
\highlight

\newcommand{\appropto}{\mathrel{\vcenter{
  \offinterlineskip\halign{\hfil$##$\cr
    \propto\cr\noalign{\kern2pt}\sim\cr\noalign{\kern-2pt}}}}}

\author{Florian Pabst\,\orcidlink{0000-0001-9331-5172}}
\email{fpabst@sissa.it}
\affiliation{%
 SISSA – Scuola Internazionale Superiore di Studi Avanzati, Trieste (Italy, EU)
}%

\author{Stefano Baroni\,\orcidlink{0000-0002-3508-6663}}%
 \affiliation{%
 SISSA – Scuola Internazionale Superiore di Studi Avanzati, Trieste (Italy, EU)
}%
\altaffiliation{CNR-IOM, Istituto dell'Officina dei Materiali, SISSA unit, Trieste (Italy, EU)}

\title[An \textsf{achemso} demo]
  {How Salt Solvation Slows Water Dynamics While Blue-Shifting Its Dielectric Spectrum}

\begin{document}







\begin{abstract}
Water inherently contains trace amounts of various salts, yet the microscopic processes by which salts influence some of its physical properties remain elusive. Notably, the mechanisms that reduce the dielectric constant of water upon salt addition are still debated. The primary absorption peak for electromagnetic radiation---commonly used in microwave heating---shifts towards higher frequencies in saline solutions, suggesting faster water molecular dynamics. This observation, however, contrasts with the simultaneous increase in viscosity and experimental reports that ionic solutes would slow down water molecular motion. In this work, we use molecular dynamics (MD) simulations with deep-neural-network models trained on high-quality quantum mechanical data to mimic interatomic forces and molecular dipoles, to compute the dielectric spectra of perchlorate water saline solution, which may be relevant to the recent discovery of liquid water beneath the thick ice crust at Mars’s south pole. Our results reveal that both the reduction in the dielectric constant and the absorption peak shift can be attributed to ion-induced changes in the orientational ordering of water molecules. Additionally, we demonstrate that the self-part of the molecular dipole-dipole correlation function reveals clear signatures of the slowing dynamics within the first cationic solvation shell, consistent with the experimentally observed increase in viscosity.
\end{abstract}

\section{Main text}

Heating salted water in a microwave oven is an everyday task. Yet, a detailed microscopic understanding of how the presence of solvated ions alters the response of the water molecules to the applied electric field is still incomplete. While such an in-depth understanding is unimportant for cooking, it is highly desirable in science, from both fundamental and applicative perspectives. In planetary sciences, for instance, the search for exo-terrestrial life is closely tied with the search for alien liquid water. In the case of Mars, for instance, data from MARSIS---a radar sounder orbiting the planet---have recently been interpreted as evidence of a liquid water body beneath a thick ice crust at the Martian south pole \cite{orosei2018radar,lauro2021multiple}. In order for water to be liquid at the presumed temperatures in this area, its freezing temperature has to be notably depressed, which is a common effect of salt. In fact, several types of salts have been found on Mars \cite{hecht2009detection}, making it plausible that the liquid water is actually brine. Since the MARSIS probe is operating only in a very narrow frequency range in the MHz band, from which the dielectric constant of the material reflecting the electromagnetic waves is estimated, it is important to understand how the dielectric spectrum evolves as a function of salt concentration. However, this matter is an open problem since more than a century \cite{bluh1924dielektrizitatskonstanten,buchner1999dielectric,lunkenheimer2017electromagnetic,galindo2023microwave}. Experimentally, two universal observations are made irrespective of the specific salt under study: With increasing salt concentration, the dielectric constant decreases monotonically, and the absorption peak in the dielectric loss shifts to higher frequencies \cite{hasted1948dielectric,buchner2009interactions}. The former phenomenon is historically ascribed to dielectric saturation, i.e. it is thought to arise because ions in salt solutions orientationally “lock” nearby water molecules with their own electric field, preventing them from aligning freely as they would in pure water.\cite{bluh1924dielektrizitatskonstanten,buchner2009interactions} Thus, a smaller number of water molecules would contribute to the dielectric response, lowering the dielectric constant. However, this picture is problematic from a dynamic point of view: While both experiments and simulations show that the rotational mobility of water molecules is hindered in the proximity of most ions,\cite{omta2003negligible,vila2013cooperative,zhang2017molecular,gonzalez2023lifting,van2023dynamics} this slowing down is not expected to affect a static susceptibility, such as $\varepsilon_s = \varepsilon'(\omega\rightarrow0)$.

Also not fitting into the picture is the shift of the dielectric loss peak to higher frequencies upon increasing salt concentration, which, as mentioned above, seems to imply \emph{faster} dynamics of the molecules, instead. Additional slow modes have occasionally been reported to affect dielectric spectra. However, they are quite weak in this case \cite{buchner2023ion}. In contrast, optical Kerr effect spectra of various salt solutions were found to exhibit evidence of intense slow water modes already at low salt concentrations \cite{gonzalez2023lifting}, an interpretation that was recently challenged\cite{zeissler2025fresh}. 

Since the classical work of Kirkwood,\cite{kirkwood1939dielectric} a key connection has been established between the dielectric constant and the orientational ordering of the molecules, as described by the Kirkwood $g_K$ factor,
\begin{equation}
     g_K = 1 + \frac{1}{N}\sum\limits_{i} \sum\limits_{j\neq i} \left<\bm{\mu}_i  \bm{\mu}_j\right>.
     \label{eq:gK}
\end{equation}
However, dielectric experiments alone cannot disentangle the effects of a decreasing $g_K$ factor from those of a decreasing number of mobile molecules contributing to the dielectric properties. Recent ab initio simulations\cite{zhang2023dissolving} showed that the decrease in $g_K$ is indeed the leading mechanism that determines the reduced static dielectric response in salt solutions. More generally, it has been shown both theoretically\cite{dejardin2019linear,dejardin2022temperature} and experimentally\cite{pabst2020dipole,pabst2021generic} that orientational correlations have a significant impact on the dielectric spectrum across a broad range of liquids. The present study aims to extend the analysis or Ref. \citenum{zhang2023dissolving} from the static to the dynamic regime, by calculating with first-principles accuracy the full dielectric spectrum for water mixed with calcium perchlorate Ca(ClO$_4$)$_2$---a salt found to be abundant on Mars\cite{hecht2009detection}---in order to elucidate the role of ionic solutes in the dielectric response of water.

To this end, we used the DeePMD kit\cite{zhang2020dp} and the DP-GEN\cite{zeng2023deepmd} iterative learning scheme to train a neural network potential (NNP) on density functional theory (DFT) data, which was obtained using \qe \cite{giannozzi2009quantum} with the RPBE-D3(BJ) exchange-correlation functional.\cite{hammer1999improved,grimme2011effect} Simulations with the NNP are run using LAMMPS\cite{thompson2022lammps} with $\approx$ 512 water molecules and the appropriate amount of ions to obtain the concentrations as indicated. All details can be found in the supporting information (SI). It was shown previously that ab-initio simulations using this level of theory are able to reproduce the experimental dielectric spectrum of pure water\cite{holzl2021dielectric}. However, the computational cost for obtaining converged spectra with this kind of fully ab-initio simulations is too high to be extended to several salt concentrations. With our NNP in combination with a second neural network trained to reproduce the molecular electric dipoles, we are able to obtain the same spectra with a similar accuracy at a computational cost that is 3 orders of magnitude lower. The complex dielectric function is calculated from the dipole dynamical correlations via the standard formula:\cite{hansen2013theory}
\begin{align}
\varepsilon' + &  i\varepsilon'' = \frac{1}{3\varepsilon_0 k_B T V} \left(\vphantom{\int\limits_0^{\infty} }\left< \bm M(0) \bm M(0) \right> \right. \label{eq:eps} \\ +   & \left.i \omega \int\limits_0^{\infty} \exp(i\omega t) \left< \bm M(t) \bm M(0) \right> \text{d}t \right) + \varepsilon_{\infty},\nonumber
\end{align}
where $T$ is the temperature, $V$ the volume and ${\bm M}=\sum_i{\bm\mu}_i$ the total dipole of the system. 

\begin{figure}[htb!]
\includegraphics[width = 0.48\textwidth]{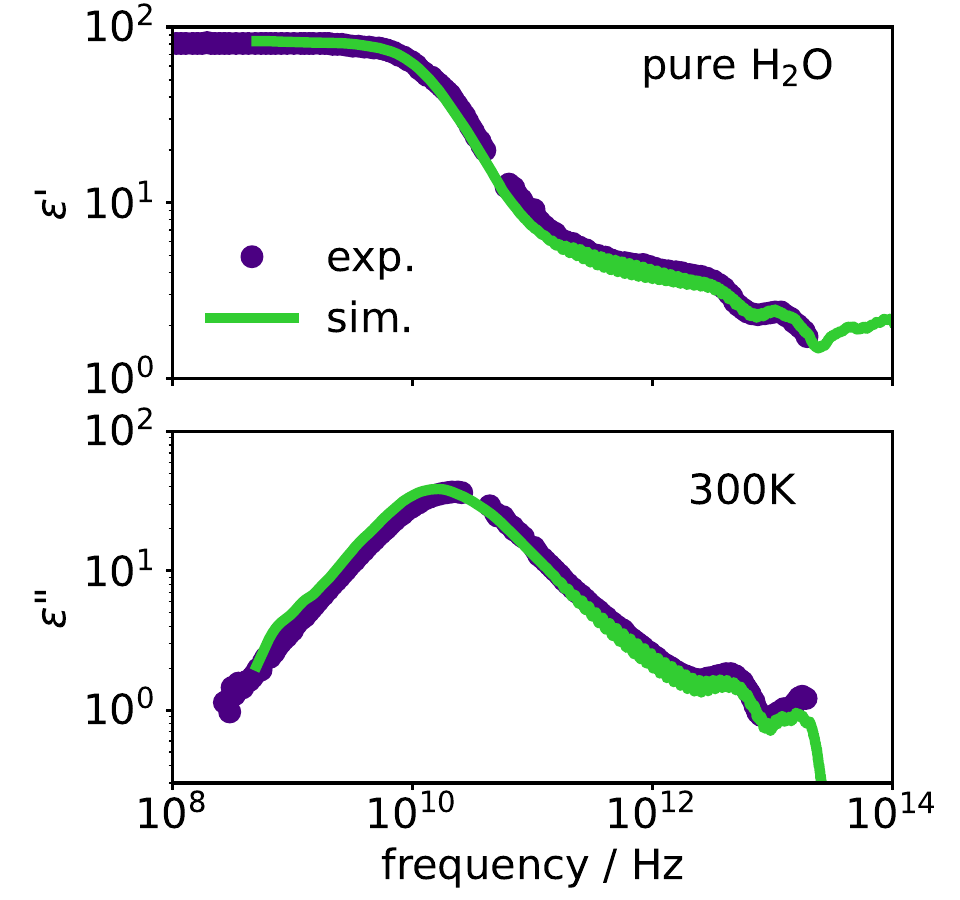}
\caption{\label{fig:exp-sim} Comparison of the results of our simulations for pure water, calculated via Eq.~\ref{eq:eps}, with experimental data\cite{lunkenheimer2017electromagnetic}. Upper panel shows the real and lower panel the imaginary part of the dielectric function.}
\end{figure}

The resulting spectrum for pure water is compared to experiments\cite{lunkenheimer2017electromagnetic} in Fig.~\ref{fig:exp-sim}, where the high frequency dielectric constant $\varepsilon_{\infty}$, not obtainable by this kind of simulation, is taken from experiments. It can be seen that the agreement is excellent, a tiny mismatch of the peak position being likely due to the temperature uncertainty of the simulation (see SI for details). 

Now that the accuracy of our approach has been benchmarked against experiments, we can move on to the salt solutions. To this end we still only consider water molecules when calculating the spectra via Eq.~\ref{eq:eps}, since the main ionic contribution to the dielectric function, eventually resulting in a finite value of the DC conductivity shows up in $\varepsilon''$ as a $\omega^{-1}$ power-law contribution, which is routinely subtracted when presenting experimental data to make the relaxation peak visible.\cite{lunkenheimer2017electromagnetic,buchner2008can} The ionic contributions to $\varepsilon'$ were found to be negligible\cite{zhang2023dissolving}, at least at lower concentrations. As concentration increase, these effects may become visible\cite{howell1974electrical,pabst2021temperature}—though this is outside the scope of the current work. 

In order to probe the effects of orientational cross-correlations on the spectra, as alluded to above, we split the spectra calculated via Eq.~\ref{eq:eps} in \emph{self} and \emph{cross} contributions, defined as:
\begin{align}
         \overbrace{\left< \bm M(t) \bm M(0) \right>}^{\rm total} = & \overbrace{\left< \sum\limits_{i=1}^{N_{\rm H_2O}} \bm\mu_i(t) \bm\mu_i(0) \right>}^{\rm self-correlation} \label{eq:total} \\ &+ \underbrace{\left< \sum\limits_{i=1}^{N_{\rm H_2O}}  \bm\mu_i(t)  \sum\limits_{i\neq j}^{N_{\rm H_2O}} \bm\mu_j(0) \right>}_{\rm cross-correlation} \nonumber
\end{align}

\begin{figure*}[ht!]
\includegraphics[width = 0.98\textwidth]{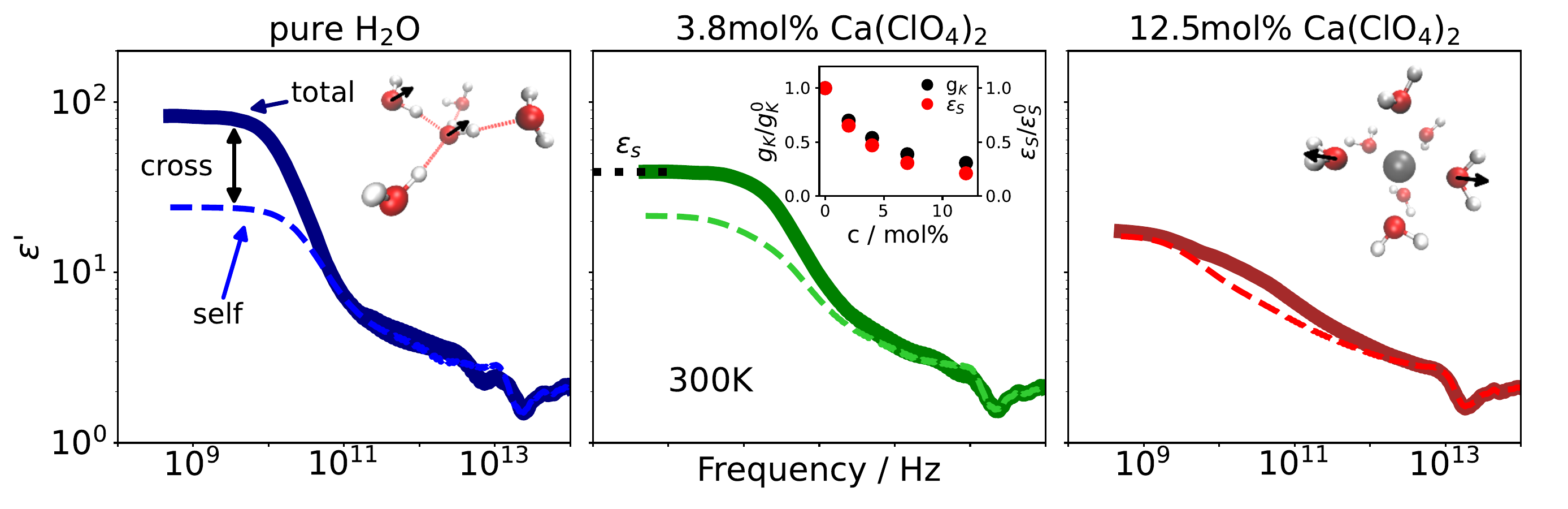}
\caption{\label{fig:eps1} Real part of the dielectric function for pure water and two salt concentrations split into total and self part (see Eq.~\ref{eq:total}). The thick continuous line indicates the total spectrum, whereas the thin dashed one indicates the self contribution. The inset in the left and right panel shows a snapshot of a water molecule hydrogen bonded to its four nearest neighbors and a calcium cation with the first solvation shell of water molecules, respectively. The inset in the middle panel show the proportionality of the Kirkwood factor $g_K$ and the dielectric constant $\varepsilon_s$ as a function of salt concentration. }
\end{figure*}

In Fig.~\ref{fig:eps1} we report the total and self contributions to the real part of the dielectric function in pure water and for two different salt concentrations. The difference between the total and self spectrum is the cross-correlation contribution (see Eq.~\ref{eq:total}). It is obvious that the plateau of the total $\varepsilon'$ at low frequencies, i.e., the static dielectric constant $\varepsilon_{\rm s}$, decreases with increasing salt concentration, and becomes closer and closer to the self contribution. This already shows that the cross-correlation contribution of water becomes less pronounced when adding salt. As mentioned above, the Kirkwood correlation factor $g_K$, defined in eq.~\ref{eq:gK}, 
is a measure of the orientational correlations between neighboring molecules. In the inset of the middle panel of Fig.~\ref{fig:eps1} the dependencies of $g_K$ and $\varepsilon_s$ on salt concentration are compared, clearly showing that the faltering angular inter-molecular correlations upon increasing salt concentration are responsible for the decrease in the static dielectric constant. The different orientational order of water molecules in pure water and around a calcium cation is shown in the two MD snapshots displayed in Fig.~\ref{fig:eps1}. While in the former case the hydrogen bond network leads to a preferred parallel orientation of neighboring dipoles, which in turn results in a $g_K$ value larger than 1 (see Eq.~\ref{eq:gK}), in the surrounding of a cation water molecules on opposite sites of the ion are oriented antiparallel to each other, leading to a reduction of the total $g_K$ value. The intensity of the self-correlation part, on the other hand, changes only marginally due to the slightly reduced number density of water molecules for higher salt concentrations. If a rotational ``locking'' mechanism would be at play, i.e., water molecules which are irrotationally bound in the first solvation shells of the ions, the self part would have to decrease significantly and be thus responsible for the decrease in $\varepsilon_s$, which is clearly not the case. 

\begin{figure}[htb!]
\includegraphics[width = 0.48\textwidth]{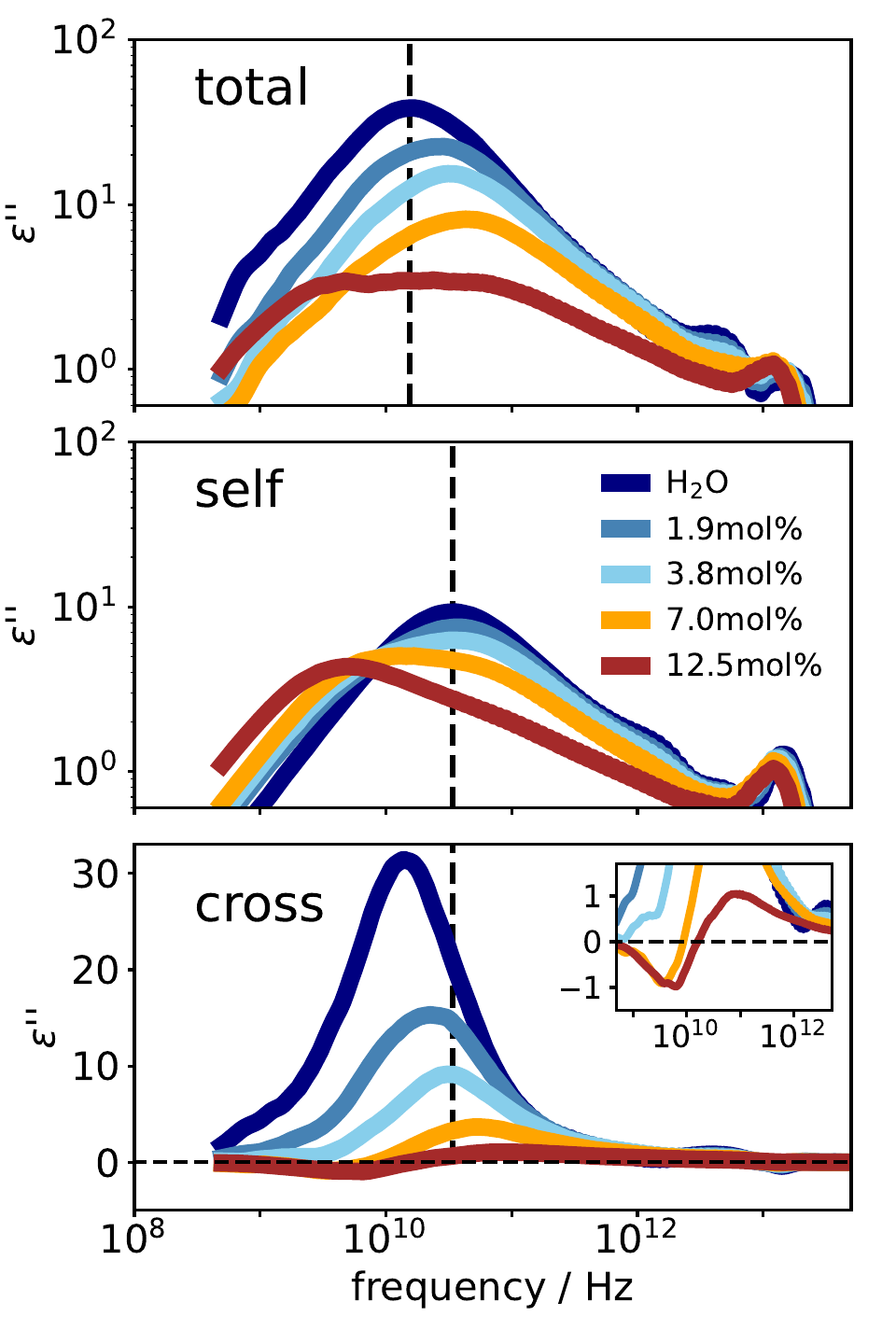}
\caption{\label{fig:eps2} Imaginary part of the total (top), self (middle) and cross (bottom) part of the dielectric functions for all concentrations studied. Please note that the y-axes of the lower panel is in linear scale in contrast to the two other panels. The vertical dashed line in the upper panel marks the peak position of neat water for the total spectrum, while the dashed vertical lines in the middle and lower part mark the peak position of the self part of neat water.}
\end{figure}

We now turn from the real to the imaginary part of the dielectric function---the dielectric loss---shown for different concentrations in Fig.~\ref{fig:eps2}. The first question we address is why the loss peak shifts to higher frequencies with increasing salt concentration---indicating faster dynamics---despite the rise in viscosity\cite{applebey1910ccxi} and results from various experimental techniques\cite{hasted1948dielectric,buchner2009interactions} suggesting the opposite.
The top panel of Fig.~\ref{fig:eps2} shows the total spectrum, once the free-ion contribution has been subtracted. A clear blue shift of the spectrum is observed with increasing concentration, eventually giving rise to a bimodal profile at the highest concentrations, to which we will return shortly. The reason for the blue shift  becomes clear when considering the self and cross parts of the spectrum, shown in the middle and lower panels of Fig.~\ref{fig:eps2}, respectively. For pure water, the peak in the self-correlation spectrum lies at a frequency approximately twice as large as in the total (and cross) spectrum (see dashed lines and note the logarithmic frequency scale), indicating that the relative orientation between neighboring molecules persists longer than the orientational correlation of individual ones.
 
Upon addition of salt, 
the intensity of the cross-correlations decreases, as previously shown, and their peak shifts to higher frequencies, causing the total spectrum to increasingly resemble its self part. At the same time, a second peak emerges on the low-frequency side of the self spectrum—initially visible only as a shoulder at low concentrations—which grows in intensity and shifts to lower frequencies as concentration increases. This slow mode first appears in the total spectrum at 7mol\% as a faint shoulder on the low-frequency side of the main peak. At the solubility limit (12.5 mol\%), the total spectrum becomes distinctly bimodal.
Altogether, these findings resolve the longstanding conundrum regarding the dielectric loss: The shift of the peak in the total spectrum to higher frequencies for higher salt concentrations is due to the decreasing intensity of the cross-correlation contribution and the assimilation of its time scale with the one of the self-correlation part. At the same time, more and more water molecules slow down their rotational motion, as manifested by the low-frequency peak in the self-part of the spectrum, which increases in intensity and shifts to lower frequencies, in line with the increasing viscosity upon salt addition. The fact that the bimodality of the total spectrum is only observed at concentrations near the solubility limit explains why it has been rarely reported in experiments before: In this case a huge DC conductivity contribution has to be subtracted from the experimental spectra, while the relaxation peak is very faint, making the subtraction procedure prone to errors or not reliable at all.  

We now address the bimodality of the spectra, which suggests the presence of molecules with two largely differing orientational correlation times. A first hint at the cause can be found by inspecting the cross-correlation spectrum of the solution with the highest salt concentration, shown enlarged in the inset of Fig.~\ref{fig:eps2}. At low frequencies, the negative peak indicates anti-correlated molecules, while the positive peak at higher frequencies reflects correlated molecules. Based on the peak position and the positive correlation, the latter can be attributed to molecules maintaining some faint orientational order, with relaxation times similar to those of pure water. In contrast, the anti-correlations seen at low frequencies likely originate from water molecules in the first cationic solvation shell of cations. As discussed above and shown in the right inset of Fig.~\ref{fig:eps1}, molecules on opposite sides of the cation are indeed orientationally anti-correlated.

\begin{figure}[htb!]
\includegraphics[width = 0.49\textwidth]{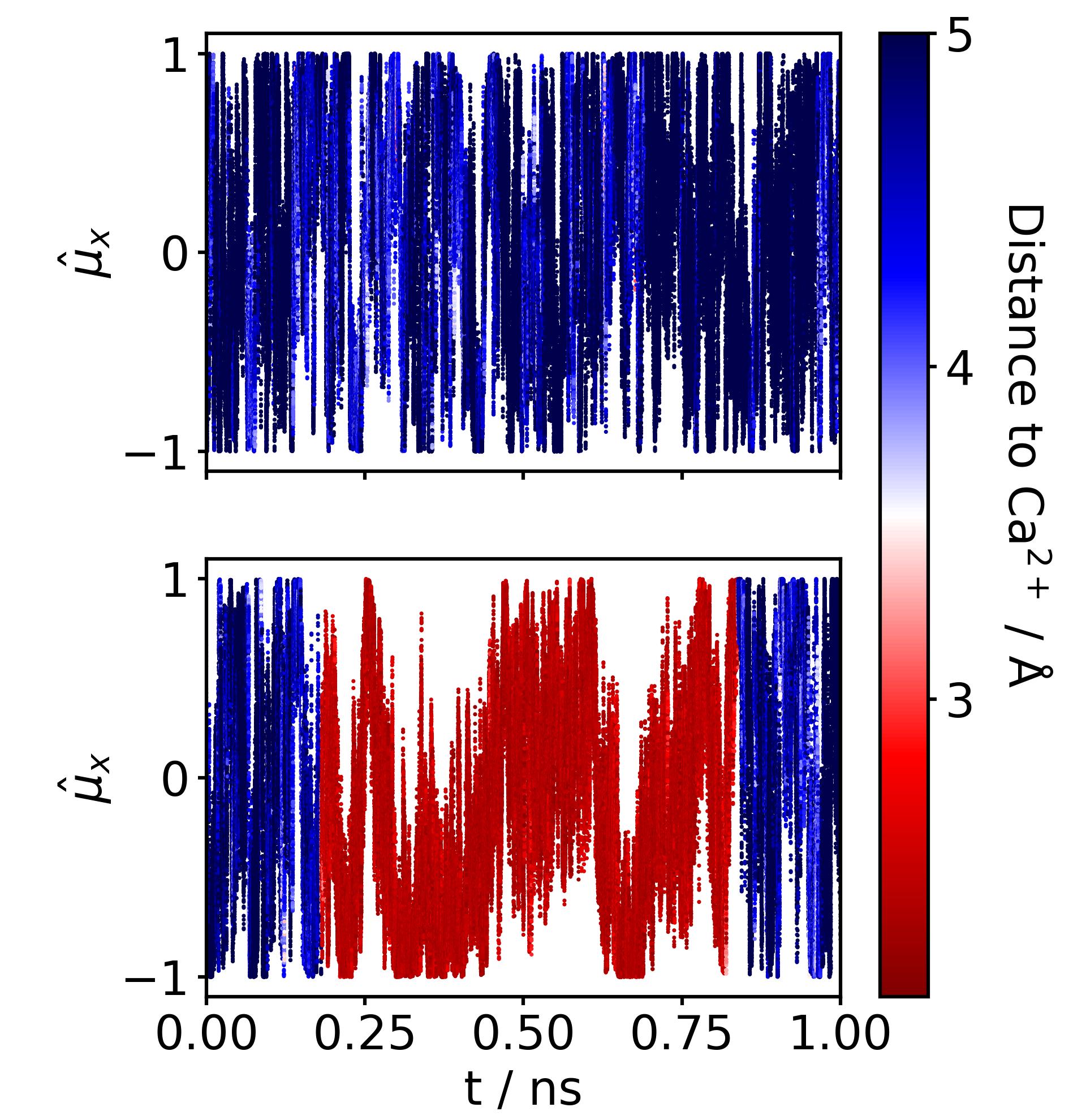}
\caption{\label{fig:dipole} Time series of the x-component of the dipole unit vector of two different water molecules (top and bottom). The color code indicates the distance to a cation.}
\end{figure}

To demonstrate that the rotational diffusion of water molecules is kinetically hindered near a solvated cation, Fig.~\ref{fig:dipole} presents a sample from the time series of the x-component of the normalized dipole of two distinct water molecules, $\hat\mu_x=\mu_x/|\bm{\mu}|$, with color indicating their distance from a reference ion. While high-frequency rotational fluctuations clearly persist as the molecule approaches the cation, it becomes evident that, upon entering the first solvation shell, a low-frequency hopping regime emerges, whereby the molecule jumps between distinct preferential orientations, with an average residence time of the order of a hundred picoseconds.

\begin{figure}[htb!]
\includegraphics[width = 0.48\textwidth]{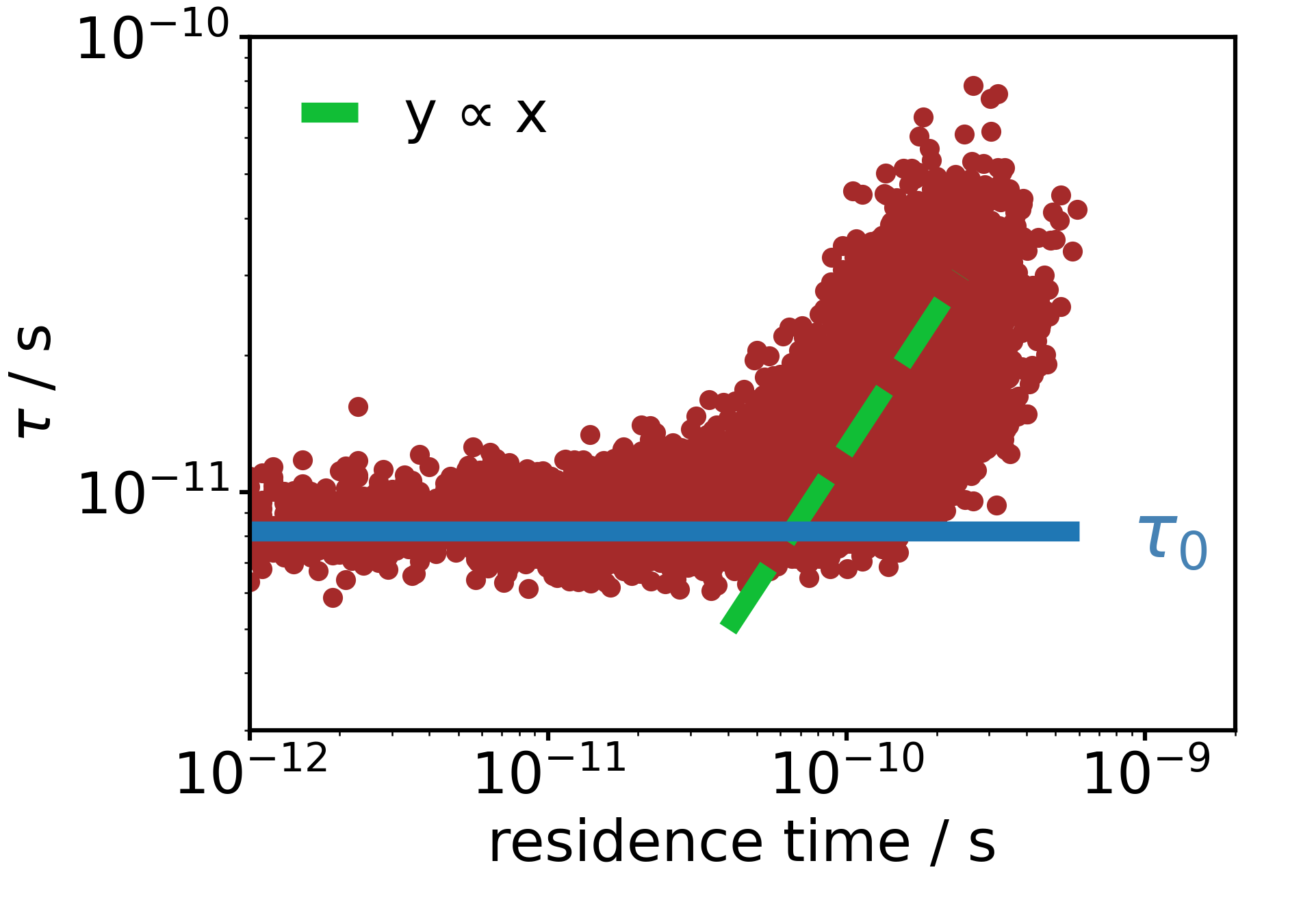}
\caption{\label{fig:contact} Dependence of the rotational correlation time of the water molecules on the residence time in a hydration shell of a cation.}
\end{figure}

In Fig.~\ref{fig:contact} the rotational correlation times $\tau$ of the water molecules in the 1.9~mol\% solution are plotted against their residence time near a cation. The latter is defined as the longest time-span within the simulation during which the molecule remains continuously in the first solvation shell of a cation. If the molecule leaves the solvation shell and comes back within a time interval of less than 200~fs, it is still considered to be continuously within the solvation shell. It can be seen that for short residence times, $\tau$ is similar to the the value of pure water, $\tau_0$, indicated by the horizontal blue line. For molecules staying longer than approximately $8\tau_0$ in the first solvation shell of a cation, a linear relationship between the residence time and the correlation time $\tau$ is observed, before a leveling off is observed after reaching the most probable residence time of around 120~ps. This clearly shows that the rotational motions of water molecules are slowed down when they are located in the first solvation shell of a cation for a significant amount of time, leading to the observed  bimodality in the dielectric spectrum.

In conclusion, our findings clarify several longstanding questions concerning the dielectric spectrum of salt solutions. The observed decrease in the dielectric constant and the shift of the main dielectric-loss peak to higher frequencies with increasing salt concentration are the equilibrium and dynamical manifestations, respectively, of a single underlying mechanism: the degradation of orientational cross-correlations among water molecules in the immediate vicinity of a solvated cation. At the same time, a 
low-frequency peak evolves in the self-part of the spectrum---in agreement with the increasing viscosity of the solution---originating from water molecules in the first hydration shell of cations which are slowed down by the presence of the cation. Thus, our results deepen the understanding of the dynamics of aqueous salt solutions. We hope that our findings will pave the way to better understand radar data for the search of liquid salty water under the surface of Mars. To this end, the next step would be to use the methodology developed in this work to calculate dielectric spectra at lower temperatures, relevant to  Martian conditions. Also, accurate thermal transport properties can be obtained in this way,\cite{tisi2021heat} which would be key to fathoming how a body of water could possibly remain liquid beneath a thick icy crust under the forbidding temperature conditions at the Martian South Pole.

\begin{acknowledgement}

This work was partially supported by the Italian MUR, through the PRIN project ARES (grant number 2022W2BPCK), by the European Commission through the \textsc{MaX} Centre of Excellence for supercomputing applications (grant number 101093374), and by the Italian National Centre for HPC, Big Data, and Quantum Computing (grant number CN00000013), funded through the \emph{Next generation EU} initiative. 

\end{acknowledgement}

\begin{suppinfo}

The supporting information contains all details on the simulations and on the training of the neural networks employed in this work, including additional references~\citenum{hamann2013optimized,lu2022dp,malosso2022viscosity,pabst2025glassy,marzari2012maximally,mostofi2008wannier90}.

\end{suppinfo}

\section*{Data Availability}
The trained neural network potential and the Wannier centroid model together with the underlying training data sets are publicly available for download at Materials Cloud \cite{pabst2025MaterialsCloud}.

\bibliography{bib}

\end{document}